\documentclass[aps,prb,twocolumn,superscriptaddress,showpacs,floatfix]{revtex4}
\usepackage{amsmath}
\usepackage{graphicx}
\usepackage{bm}

\begin{document}

\title{Can buckling account for the features seen in graphite's Raman spectra?}

\author{Y. Fujioka\footnote[1]{To whom correspondence should be addressed. E-mail: yfu@fyslab.hut.fi}
\footnote[2]{Present address: Laboratory of Physics, Helsinki University of Technology, P.O. Box 1100, 
FIN-02015 HUT, Finland.}}
\author{J. Frantti$^\dag$}
\author{E. Yasuda}
\author{Y. Tanabe}
\affiliation{Materials and Structures Laboratory, Tokyo Institute of
Technology, 4259 Nagatsuta, Midori-ku, Yokohama, 226-8503, Japan}
\author{M. Kakihana}
\affiliation{Institute of Multidisciplinary Research for Advanced Materials, 
Tohoku University, Sendai 980-8577, Japan}

\begin{abstract}
Raman scattering data were collected on graphite monochromator. Spectra were interpreted 
in terms of the space group $P6_3mc$, a subgroup of space group $P6_3/mmc$. The latter has 
commonly been used for the interpretation of Raman scattering data. Space group $P6_3mc$ 
corresponds to the buckling of graphene sheets and is consistent with many spectral features. 
Both the first and second order scattering were considered. Many first order results (most 
notably the assignments of the band at 1350 and the peak at 1620 cm$^{-1}$) were found to 
agree with previous observations [Y. Kawashima and G. Katagiri, Phys. Rev. B 
\textbf{66}, 104109 (2002)] carried out on highly oriented pyrolytic graphite samples. 
To check the consistency of the model, symmetry analysis was applied to the second order 
spectra. 
Also a simple test for buckling model was done.
\end{abstract}

\maketitle

\section{Introduction}
Despite its nominally simple crystal structure the interpretation of the Raman and 
infrared spectra of graphite has been an active and controversial topic.
The crystal symmetry of graphite is reported to be either $P6_3/mmc$ (No. 194) or 
$P6_3mc$ (No. 186) \cite{Wyckoff}. A more rare form, with rhombohedral space group 
symmetry $R\bar{3}m$ (No. 166) was also reported in ref. \onlinecite{Wyckoff}. The symmetries 
of these different forms increase in order $R\bar{3}m \rightarrow P6_3mc \rightarrow P6_3/mmc$: 
these symmetry groups share group-subgroup relationships. It is well known that, despite 
its virtually simple structure, a large number of features which seem to contradict the 
space group symmetry $P6_3/mmc$, have been observed in Raman spectra \cite{Reich}. Partially the 
problem has been that rather different types of 'graphites' were studied, as far as 
crystallite size and defects are considered. Despite that Raman spectra were commonly 
interpreted in terms of space group  $P6_3/mmc$. 

Still, the puzzling features of the Raman spectra of graphite are (i) the peak(s) at around 
1355 cm$^{-1}$ (often labelled by $D$), (ii) peak at around 1620 cm$^{-1}$ (often labelled 
by $D'$) (iii) the first overtones of the aforementioned modes (often labelled by $2D$ and 
$2D'$, respectively). If  $P6_3/mmc$ symmetry is assigned to graphite, the Brillouin zone 
centre modes transform as the irreducible representation (IRREP)
$2A_{2u}\oplus2B_{2g}\oplus 2E_{1u} \oplus 2 E_{2g}$, 
where $A_{2u}$ and $E_{1u}$ representations are infrared active, $B_{2g}$ is silent and $E_{2g}$ 
is the only Raman active representation. Thus, only two peaks, observable at $xx$, $yy$, $xy$ and $yx$ 
geometries, are expected (i.e., $E_{2g}$ modes can only be observed if the polarization vectors of both
the incoming and scattered light are perpendicular to the hexagonal $c$ axis). They were reported to 
appear at 42 and 1580 cm$^{-1}$ (the latter is commonly labelled as 'G' mode) \cite{Sinha}. However, 
the number of observed peak is generally greater. Now, it is interesting to note that in the case of 
$P6_3mc$ symmetry the Brillouin zone centre modes transform as the IRREP 
$2A_1\oplus2B_1\oplus 2E_1 \oplus 2 E_2$, where $A_1$ and $E_1$ symmetry modes are Raman and infrared 
active, $B_1$ mode is silent and $E_2$ mode is Raman active.

The first attempt to explain the $D$\mbox{ }band were based on disorder, which was interpreted to 
cause an appearance of a totally symmetric mode \cite{Tuinstra}. As is discussed below, this mode(s) 
is not totally symmetric and its origin is still not clear.
Recent explanations are based on double resonance idea \cite{Thomsen,Reich,Saito}. The 
merits of the double resonance model are its ability to take the laser beam wavelength 
dependent positions of $D$\mbox{ }and $2D$\mbox{ }bands into account. 
The first attempt to explain the $D$ band were based on an idea that, due to the disorder, 
a Raman inactive mode became active \cite{Tuinstra}. The intensity of this mode was reported 
to be dependent on particle size \cite{Tuinstra}. In addition, the frequency of this mode was reported 
to increase with increasing laser light frequency \cite{Vidano}. The corresponding changes 
occurred in the first overtone of the $D$ band. To explain these features a double resonance 
Raman scattering model was developed \cite{Thomsen,Reich,Saito}. Several weak peaks 
were observed in Raman spectra, the most notable at 867 cm$^{-1}$, which was reported to 
correspond to the out-of-plane 
vibration \cite{Kawashima1,Kawashima2}.
This model assumes that the $D$ and $D'$ 
bands are transverse optic Brillouin zone boundary ($K$ point) and zone centre near Brillouin 
zone centre modes. The double resonance model further necessitates that the atom displacements 
occur in graphene sheets (in-plane vibrations) \cite{Thomsen}.
Occasionally, double resonance 
theory has been merited to take the slight difference in the anti-Stokes and Stokes positions into 
account. In this context once should bear in mind that, in order to observe the high frequency 
overtones in anti-Stokes spectra, exceptionally high laser beam power densities were used in 
these studies (the highest powder densities were approximately 5000 times larger than the one 
used in this study) \cite{Tan1}. Although this type of considerations are interesting, we 
note that, in the case of the anti-Stokes spectra, the reported differences between the Stokes 
and corresponding anti-Stokes frequencies were, for different modes, positive, almost zero and 
negative, as was pointed out in ref. \onlinecite{Tan1}.

The purpose of this work was to reinvestigate a high quality graphite sample by Raman scattering 
technique. The first goal was to check if the buckling model can explain the 
first and second order spectra. The second goal was to consider to which extent the proposed 
double resonance model is consistent with the observed symmetry properties of $D$ and $D'$ bands.

\section{Experimental}
A pyrolytic graphite monochromator (Advanced Ceramics) sample was used in this study.
Raman measurements were performed using a Jobin-Yvon T64000
spectrometer consisting of a double monochromator coupled to the third
monochromator stage with 1800 grooves per millimeter grating 
(double substractive mode). Prior to measurements, spectrometer was 
calibrated with Ne lamp.  
Acquisition time was adjusted to have a sufficient
signal-to-noise-ratio. CCD detector was used to count  photons.
 Backscattering measurements were carried out under
the microscope. Also 90-degree scattering angle experiments (the angle between the incoming 
and scattered light was 90 degrees) were carried out. Raman spectra were excited using an argon ion laser. 
In the case of backscattering experiments, the laser beam power on the sample surface
 was 200 $\mu$W and the diameter of the laser beam spot was 
approximately 2 $\mu$m (the spot diameter was approximately 100 $\mu$m in 90-degree scattering angle 
measurements, and the power was 50 mW, so that the powder density was smaller than in the 
case of backscattering measurements). In a backscattering geometry the wavelength was 
514.532 nm and in a $90$-degree scattering angle measurements it was 487.986 nm.

\section{Results and discussion}

\begin{figure*}[hbt!]
\begin{center}
\includegraphics[width=11.0cm,angle=0]{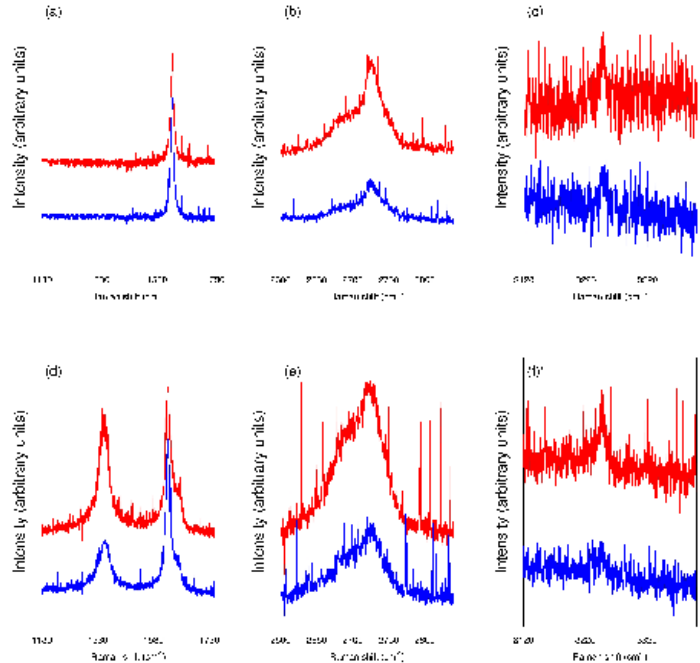}
\caption{\label{Fig1} Raman spectra collected on $ab$ plane at $Z(XX)\bar{Z}$ and $Z(XY)\bar{Z}$ 
geometries: (a) spectral region corresponding to the $D$, $G$ and $D'$ modes, (b) Spectral region 
corresponding to the $2D$ mode(s) and (c) $2D'$ modes. The corresponding Raman spectra collected 
on edge plane at $X(ZZ)\bar{X}$ and $X(YZ)\bar{X}$ geometries are shown in panels (d), (e) and (f). 
Spectra shown in red correspond to the setting where polarizer and polarizator were parallel to 
each other, whereas blue lines give the spectra collected on with polarizer perpendicular to the 
incoming polarization direction.} 
\end{center}
\end{figure*}

Selected regions of Raman spectra collected on $ab$-basal plane with an polarizer set parallel 
(labelled as $Z(XX)\bar{Z}$) and perpendicular (labelled as $Z(XY)\bar{Z}$) with respect 
to the incoming polarization direction are shown in Fig. \ref{Fig1} (a), (b) and (c). Particular 
attention was paid on the $D$, $D'$, $G$, $2D$, and $2D'$ modes. Here $Z$ and $\bar{Z}$ 
refer to the incoming and scattered light propagation direction. Although the light propagation 
direction can be rather accurately determined at backscattering geometries, incoming light 
polarization direction with respect to hexagonal $a$ and $b$ axes was not known. The corresponding 
regions, obtained through right-angle geometry, are shown in Fig. \ref{Fig1} (d), (e) and (f).
The intensity of $G$ mode at around 1580 
cm$^{-1}$ was the same for $XX$ and $XY$ geometries (see Fig. \ref{Fig1} (a)), consistently with the idea that it belongs 
to the $E_{2g}$ symmetry (space group $P6_3/mmc$) or $E_2$ symmetry (space group $P6_3mc$). In 
contrast to the earlier reports (see, e.g., ref. \onlinecite{Sinha}), we could not see any other mode below 
1580 cm$^{-1}$. Fig. \ref{Fig1} (b) shows the broad band centered at around 2750 cm$^{-1}$. We 
note that this band is consisted of several peaks and the intensity differs 
from zero in $XX$, $XY$, $ZZ$ and $ZY$ geometries. 
Now, it is of interest to consider the symmetries of the first overtones (second order scattering) 
in the case of the fundamentals belonging to the point group symmetries $6/mmm$ and $6mm$. 
Only two-dimensional IRREPs need to be considered, since the first overtone of a one dimensional 
IRREP is always totally symmetric and can be distinguished by depolarization measurements. 
Now, the first overtone of the $E_{1u}$, $E_{1g}$, $E_{2u}$ and $E_{2g}$ modes transform as 
$A_{1g}\oplus E_{2g}$ and the first overtone of the $E_1$\mbox{ }mode transforms as 
$A_1\oplus E_2$. Thus, if this band is due to the second order scattering 
of the Brillouin zone centre mode, it must belong to $E_{1u}$ (space group $P6_3/mmc$) 
or to $E_1$ symmetry (space group $P6_3mc$).  To check if either assignment is correct two 
other experimental geometries were used. 
First one is the geometry where 
light propagates perpendicularly to the hexagonal $c$-axis, Fig. \ref{Fig1} (d)-(f). Consistently, $G$ 
mode was significantly weaker (if the polarization direction of the incoming or scattered light is 
strictly parallel to the hexagonal axis, the $G$ mode should have zero intensity. Deviations might 
be because of the folding of graphene sheets). The appearance of the mode at around 1355 cm$^{-1}$ is 
consistent with the space group $P6_3mc$. Basically, if this mode belongs to the $E_1$ symmetry, 
it should only be observed at the $XZ$\mbox{  }and $YZ$ geometries. Thus, this mode should not be observable 
at $ZZ$ geometry, in constrast to the experimental observations. Although partially this might be 
due to the aforementioned folding we decided to carry out 90-degree measurements. 

\begin{figure}[hbt!]
\begin{center}
\includegraphics[width=8.6cm,angle=0]{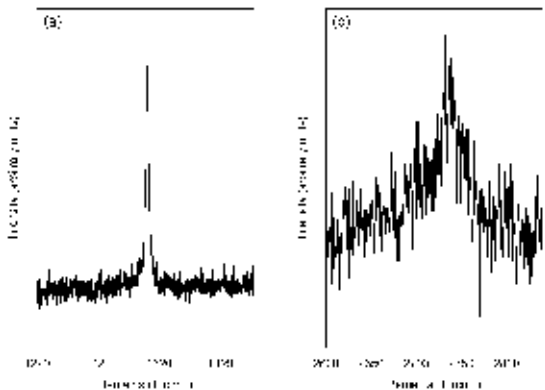}
\caption{\label{Fig2} Raman spectra collected on $90$-degree scattering angle geometry, where phonon 
propagates along the graphene sheet. (a) spectral region between which should reveal also the $D$ 
and $D'$ modes, (b) Confirmation that $2D$ are observed at this geometry, too.} 
\end{center}
\end{figure}

If the buckling model alone is sufficient for explaining the appearance of $D$ and $D'$\mbox{ }bands 
(instead of the disorder, which activates non-Brillouin zone centre modes), one should see them 
at 90-degree measurements carried out on large graphene sheets (so that the effect of disorder 
as a dominant factor can be eliminated). Now, the phonon propagates parallel to the graphene sheet, 
by momentum conservation rule. The $G$ mode was observed (Fig. \ref{Fig2}), since both the incoming 
and right-angle scattered light have a polarization component parallel to the $ab$ plane. The 
more interesting point is that now also the modes belonging to the $E_1$ symmetry modes should 
be observed. Thus, this experiment served as a test to clarify whether $D$\mbox{ }and/or $D'$ modes 
are due to the Brillouin zone centre modes or if they correspond to the disorder activated modes.
As Fig. \ref{Fig2} reveals, $D$\mbox{ }and $D'$ bands were not observed, but $2D$ was observed.
Rather common opinion is that $D$ mode is related to the Brillouin zone $K$ point mode (sometimes 
assigned to the totally symmetric mode), whereas $D'$\mbox{ }mode has sometimes been considered to belong 
to $B_1$ symmetry with wavevector close to zero (Brillouin zone centre), see for example ref. \onlinecite{Reich}.
Now, the problem is that these assignments do not comfort with the polarization measurements. The first 
overtone of these modes should also be totally symmetric, in contradiction with the experiments. Now, the 
selection rules for phonons which can participate on double resonance scattering state that only totally 
symmetric modes can resonantly couple electrons within the same non-degenerate bands \cite{Reich}. 
The second possibility is that the phonon involved in the process should belong to the $B_1$ symmetry. 

\section{Conclusions}
The Raman spectra collected on graphite monochromator were interpreted in terms of two space group symmetries.
The only observed first order lines were $G$ mode, and $D$ and $D'$ modes. It was concluded that although the 
weak, additional lines reported in literature probably do correspond to the buckling of graphene sheets, 
the $D$ and $D'$ modes are probably not due to the buckling. The latter feature was confirmed by $90$-degree 
angle experiments. However, the situation is different for smaller crystal size materials. It was also confirmed 
that $D$ and $D'$ modes do not belong to the totally symmetric representation. 

%\begin{table}
%\begin{center}
%\caption{\label{Fundamentals} The symmetries of the first overtones (second order scattering) 
%in the case of the fundamentals belonging to the point group symmetries $6/mmm$ and $6mm$. 
%Only two-dimensional IRREPs were considered, since the first overtone of a one dimensional 
%IRREP is always totally symmetric and can be distinguished by depolarization measurements.} 
%\begin{tabular}{l l l l l}
%\hline
%$6/mmm$     &                     &          &          &          \\
%Fundamental & $E_{1u}$            & $E_{1g}$ & $E_{2u}$ & $E_{2g}$ \\
%Overtone    & $A_1g\oplus E_{2g}$ &          &          &          \\
%\hline
%$6mm$       &                     &          &          &          \\
%Fundamental & $E_1$               &          & $E_2$    &          \\
%Overtone    &                     &          &          &          \\
%\hline
%\end{tabular}
%\end{center}
%\end{table}

\end{document}